\documentclass[aps,twocolumn,10pt,showpacs,superscriptaddress,amsmath,amssymb]{revtex4-1}
\usepackage{graphicx}
\usepackage{amsmath}  
\usepackage{amsfonts} 
\usepackage{dcolumn}
\usepackage{bm}
\usepackage[T1]{fontenc}
\usepackage{color}

\begin{document}

\title{A connection between non-local one-body and local three-body correlations of the Lieb-Liniger model}

\author{Maxim Olshanii} 
\affiliation{Department of Physics, University of Massachusetts Boston, Boston, MA 02125, USA}
\email{Maxim.Olchanyi@umb.edu}

\author{Vanja Dunjko} 
\affiliation{Department of Physics, University of Massachusetts Boston, Boston, MA 02125, USA}

\author{Anna Minguzzi}
\affiliation{Universit\'e Grenoble-Alpes, CNRS, LPMMC  F-38000 Grenoble, France}

\author{Guillaume Lang}
\affiliation{Universit\'e Grenoble-Alpes, CNRS, LPMMC  F-38000 Grenoble, France}


%
\begin{abstract}
We derive a connection between the fourth coefficient of the short-distance Taylor expansion of the one-body correlation function, and the local three-body correlation function of the Lieb-Liniger model of $\delta$-interacting spinless bosons in one dimension. This connection, valid at arbitrary interaction strength, involves the fourth moment of the density of quasi-momenta. Generalizing recent conjectures, we propose approximate analytical expressions for the fourth coefficient covering the whole range of repulsive interactions, validated by comparison with accurate numerics. In particular, we find that the fourth coefficient  changes sign  at interaction strength $\gamma_c\simeq 3.816$, while the first three coefficients of the Taylor expansion of the one-body correlation function retain 
the same sign throughout the whole range of interaction strengths.

%
%
\end{abstract}

\maketitle


\section{Introduction}

One-dimensional quantum systems are generically more strongly correlated than their higher-dimensional counterparts, due to the inevitability of collisions when particles cross each other.  The one-dimensional Bose gas with contact interactions, known as the Lieb-Liniger model, is the paradigm of such systems \cite{Lieb}. This model well describes experiments with ultracold atoms in tight waveguides  and some of its correlation functions have been experimentally probed in all interaction regimes \cite{Paredes, Kinoshita, Kinoshitabis, Haller, Hallerbis, Bouchoule, Clement, Meinert}. From a theoretical point of view, since the model is integrable, its $k$-body correlations can in principle be obtained explicitly at all orders $k$, since they are linked to the (infinite) set of integrals of motion. In practice, however, the exact analytical calculation at arbitrary interaction strength is tremendously difficult for two reasons: the coefficients of their Taylor expansion at small distance are related to each other in a non-trivial way, and the defining system of equations, in turn derived using Bethe Ansatz, 
is technically very challenging. In the thermodynamic limit, finding an explicit expression for the various spatially 
non-local, equal time correlations from Bethe Ansatz techniques actually requires to link them to the moments of the density of quasi-momenta, and thus to solve a type II homogeneous Fredholm integral equation with Lorentzian kernel, whose exact analytic solution is yet unknown.

In this work, we focus on the link between non-local correlation functions of the Lieb-Liniger model and its integrals of motion, thus elucidating a special structure of the ground state for this integrable model. In particular, we derive a relation, first proposed in \cite{Dunjko}, that links the fourth coefficient of the Taylor expansion of the one-body correlation function at short distances with  various moments of the quasi-momentum distribution and their derivatives with respect to the coupling constant. Then, we use a recently-developed method \cite{Ristivojevic, Lang2016}, and generalize recent conjectures \cite{Widom, Lang2016, Prolhac}, to evaluate these quantities with excellent accuracy in a wide range of interaction strengths.

The paper is organized as follows: in Section \ref{Hamilton} we introduce the Hamiltonian of the system and the relevant notations. We also define the notion of connection, which is the key concept in this work, and illustrate it on simple cases. Then, in Sec.~\ref{Connect} we derive a connection between the fourth coefficient of the Taylor expansion of the one-body correlation function at short distances, $c_4$, the local three-body correlation function and the fourth moment of the density of pseudo-momenta, which is one of the main results of this work. In Section \ref{Conjecturesmoments}, we provide new conjectures about the moments of the density of pseudo-momenta, and illustrate them in Sec.~\ref{Illus} where we find, in particular, that $c_4$ changes sign at interaction strength $\gamma\simeq 3.8$. In Section \ref{Outlook}, we summarize our main results and give an  outlook to our work.

\section{Hamiltonian and definitions}
\label{Hamilton}

In this work, we consider a one-dimensional system of $N$ indistinguishable point-like bosons of mass $m$ subject to contact interactions, known as the Lieb-Liniger model. We choose periodic boundary conditions, possibly realized by using a ring geometry.
The Hamiltonian of the system reads
\begin{align}
\hat{\cal H} =\frac{\hbar^2}{2m} \left[
 \sum_{i=1}^{N} -\frac{\partial^2}{\partial x_{i}^2}
+  2 c \sum_{i=1}^{N-1} \sum_{j=i+1}^{N} \delta(x_{i}-x_{j})\right]
\,\,,
\label{H}
\end {align}
where the first term of the right-hand side stands for the kinetic energy, $\{x_i\}$ label the positions of the atoms, $\delta$ is the Dirac delta function, $c\!=\!2/a$ is related to  the coupling constant, with $a\!=\!-a_{\mbox{\scriptsize 1D}}\!>\!0$ for repulsive interactions, and  $a_{\mbox{\scriptsize 1D}}$ is the one-dimensional
scattering length related to the many-body wavefunction by $\Psi(\ldots,\,x_{i},\,\ldots ,x_{j},\,\ldots) \propto |x_{i}\!-\!x_{j}|-a_{\mbox{\scriptsize 1D}}+\ldots$. The usual one-dimensional coupling constant of the model is $g_{\mbox{\scriptsize 1D}}\!=\!-2\hbar^2/(m a_{\mbox{\scriptsize 1D}})\!=\!(\hbar^2/m)c$ \cite{Olshaniibis}.

The action of the Hamilonian on the Bethe ground state $|\chi_N\rangle$ of the many-body system reads
\begin{align}
\hat{\cal H}|\chi_N \rangle = \sum_{i=1}^{N} \lambda_{i}^2 |\chi_N \rangle
\,\,,
\label{H_2}
\end {align}
the eigenvalue is the sum of Bethe rapidities $\lambda_{i}$ squared, and the ground state in coordinate representation reads \cite{Gaudin, Korepin}
\begin{widetext}
\begin{align}
\chi_{N}(x_{1},\,\ldots,x_{N}) = \mbox{const} \times\!\! 
\sum_{\sigma \in S_N} (-1)^{{\cal P}[\sigma]}\prod_{i=1}^{N-1}\prod_{j=i+1}^{N} 
\left[\lambda_{\sigma_{i}} - \lambda_{\sigma_{j}} - \frac{2 i}{a} \mbox{sign}(x_{i}-x_{j})\right] \exp\left[i \sum_{k=1}^{N} \lambda_{\sigma_{k}} x_{k}\right].
\label{chi}
\end {align}
\end{widetext}
In Eq.~(\ref{chi}), $\sigma$ are elements of the symmetry group $S_N$, i.e. permutations of $N$ elements, ${\cal P}[\sigma]$ their parity, and $\sigma_i$ denotes the image of $i$ by $\sigma$. These eigenstates are also eigenfunctions of all conservation laws, as required by the integrability of the model.

The correlation functions we are interested in are the $k$-body density matrices normalized to unity by choice of the constant in Eq.~(\ref{chi}), and such that
\begin{align}
&\rho_k(x_1,\dots x_k;x_1',\dots, x_k')\equiv \int\!dx_{k+1}\dots dx_N \nonumber\\
&\chi_N^*(x_1',\dots, x_k',x_{k+1},\dots,x_N)\chi_N(x_1,\dots,x_N)\nonumber\\
&=\rho_k(x_1-x_1',\dots, x_k-x_k';0,\dots, 0)
\end{align}
due to Galilean invariance. In particular, in this work we consider the one-body density matrix, whose series expansion at short distance can be written as
\begin{align}
&
\rho_{1}(x;\,x') =\frac{1}{L}\sum_{l=0}^{+\infty}c_l(n|x\!-\!x'|)^l.
\end{align}

In what follows we will also make use of the notation
\begin{eqnarray}
g_k\equiv \frac{N!}{(N-k)!}\frac{\rho_k(0, \dots, 0 ; 0 ,\dots, 0)}{n^k},
\end{eqnarray}
where $n=N/L$ is the mean linear density, the system being of size $L$. Correlations $g_k$ will be refered to as $k$-body local correlations, they represent the probability to find $k$ atoms at the same place and time. It is quite intuitive that the combined effect of geometry and interactions enforces $g_{k+1}<g_k$ at finite interaction strengths, and that $g_k$ are decreasing functions of the interaction strength.

The first aim of this work is to illustrate the fact that, due to the integrability of the model, those correlations are related to each other via the moments $e_{2k}$ of the dimensionless density of quasi-momenta $g(z;\alpha)$, defined as the solutions of the set of Bethe equations derived by Lieb and Liniger \cite{Lieb} in the thermodynamic limit $N\to \infty$, $L\to \infty$, at fixed $N/L$ and at zero temperature:
\begin{equation}
\label{Fredholm}
g(z;\alpha)-\frac{1}{2\pi}\int_{-1}^{1}dy\frac{2\alpha g(y;\alpha)}{\alpha^2+(y-z)^2}=\frac{1}{2\pi},
\end{equation}
\begin{equation}
\label{alphagamma}
\gamma \int_{-1}^1dy g(y;\alpha)=\alpha,
\end{equation}
and
\begin{equation}
\label{moments}
 e_{2k}(\gamma)=\frac{\int_{-1}^1dy g(y;\alpha(\gamma))y^{2k}}{[\int_{-1}^1 dy g(y;\alpha(\gamma))]^{2k+1}},
\end{equation}
where $\alpha$ is a positive coefficient and $\gamma\!\equiv\!2/(na)$ is the Lieb parameter, representing the natural dimensionless coupling constant of the model.
The quantities $e_{2k}$ are integrals of motions of the model: for exemple, $e_2$ corresponds to the thermodynamic limit of the ground-state energy $E_2=\sum_i \lambda_i^2$, through $E_2=Nn^2e_2$. More generally, we define $E_{2k}= Nn^{2k}e_{2k}$, and $e_{2k+1}=0$ from parity arguments.

To finish with, we define connections as functionals $\mathcal{F}$ such that
\begin{eqnarray}
\mathcal{F}\left(c_l(\gamma),g_k(\gamma),\{e_{n}(\gamma),e_n'(\gamma),\dots\},\gamma\right)=0,
\end{eqnarray}
where $'$ denotes differentiation with respect to $\gamma$. We denote each connection by a pair of indices $(l,k)$, where by convention an index is $0$ if the corresponding quantity in the notation above does not appear in the functional. This compact notation helps classifying the connections. To illustrate this concept, we derive the first few connections from conservation laws.

The first conserved quantity is
\begin{eqnarray}
\hat{H}_0=\sum_{i=0}^N\frac{\partial^0}{\partial x_i^0}=N,
\end{eqnarray}
the number of atoms.
Trivially,
\begin{eqnarray}
\langle \chi_N|\hat{H}_0|\chi_N\rangle=N\langle \chi_N|\chi_N\rangle=N
\end{eqnarray}
since the Bethe eigenstate is normalized to unity. On the other hand,
\begin{eqnarray}
&&\langle \chi_N|\hat{H}_0|\chi_N\rangle\nonumber\\
&&=\!N\!\int\!dx_1\dots dx_N\chi_N^*(x_1,\dots,x_N)\chi_N(x_1,\dots,x_N)\nonumber\\
&&=N\int dx_1\int dx_1' \delta(x_1-x_1')\int dx_2\dots dx_N\nonumber\\
&&\chi_N^*(x_1',\dots,x_N)\chi_N(x_1,\dots,x_N)\nonumber\\
&&=NL\rho_1(x;0)|_{x=0}=Nc_0=Ng_1,
\end{eqnarray}
hence
\begin{eqnarray}
c_0=g_1=e_0=1,
\end{eqnarray}
yielding the connections (0,0) and (0,1).

The second conserved quantity is
\begin{eqnarray}
\hat{H}_1=\sum_{i=0}^N\frac{\partial}{\partial x_i}
\end{eqnarray}
and proceeding as before, we find
\begin{eqnarray}
c_1=e_1=0,
\end{eqnarray}
the connection of type (1,0), in agreement with \cite{Olshanii}.

Then, from the Hamiltonian we obtain
\begin{eqnarray}
\langle\chi_N|\hat{\cal H}|\chi_N\rangle=Nn^2e_2.
\end{eqnarray}
We also evaluate
\begin{eqnarray}
&&\langle \chi_N|\sum_{i=1}^N-\frac{\partial^2}{\partial x_i^2}|\chi_N\rangle=NL\frac{\partial^2}{\partial x^2}\rho_1(x;0)|_{x=0}\nonumber\\
&&=2Nn^2c_2,
\end{eqnarray}
and
\begin{eqnarray}
\langle \chi_N|\sum_{i=1}^{N-1} \sum_{j=i+1}^{N} \delta(x_{i}-x_{j})|\chi_N\rangle=\frac{N}{2}ng_2
\end{eqnarray}
hence the connection of order (2,2),
\begin{eqnarray}
-2c_2+\gamma g_2=e_2.
\end{eqnarray}
The connection of type (0,2) is obtained by applying the Hellmann-Feynman theorem to the Hamiltonian and reads \cite{Ganshlya}
\begin{eqnarray}
\label{Eq02}
g_2=e_2',
\end{eqnarray}
combining the connections of orders (2,2) and (0,2) yields order (2,0), i.e. \cite{Olshanii}
\begin{eqnarray}
c_2=\frac{1}{2}(\gamma e_2'-e_2).
\end{eqnarray}

\section{Derivation of the connection of order $(4,3)$}
\label{Connect}
In this Section, we derive a new connection, namely

\begin{align}
24 c_{4} - 2 \gamma^2 g_{3} = e_{4}-\gamma e_{4}'.
\label{Maineq}
\end{align}
It is the connection of type (4,3) according to our nomenclature.

%

%
%

First, we introduce an operator $\hat{H}_{4}$ that yields, when applied to an eigenstate (\ref{chi}), the fourth integral of motion $E_4$,

\begin{align}
\hat{H}_{4} | \chi_N \rangle = E_{4} | \chi_N \rangle
\,\,,
\end {align}
with 
\begin{align}
E_{4}= \sum_{i=1}^{N} \lambda_{i}^4.
\,\,
\end {align}
From these definitions, by construction the higher Hamiltonian $\hat{H}_4$ can be written explicitly as \cite{Gutkin, KorepinDavies, Davies}
\begin{align}
%
%
&\hat{H}_{4}= \sum_{i=1}^{N} \frac{\partial^4}{\partial x_{i}^4}
+
\frac{48}{a^2} \sum_{i=1}^{N-2} \sum_{j=i+1}^{N-1}\sum_{k=j+1}^{N} \delta(x_{i}-x_{j}) \delta(x_{j}-x_{k})\nonumber
\\
&
\quad
- \frac{4}{a}  \sum_{i=1}^{N-1} \sum_{j=i+1}^{N} 
\left\{
   \left(
    \frac{\partial^2}{\partial x_{i}^2} + \frac{\partial^2}{\partial x_{j}^2} + \frac{\partial^2}{\partial x_{i} \partial x_{j}}
    \right)\delta(x_{i}-x_{j})
\right.\nonumber
\\
&
\qquad\quad
\left.
    +
     \delta(x_{i}-x_{j})  
     \left(
     \frac{\partial^2}{\partial x_{i}^2} + \frac{\partial^2}{\partial x_{j}^2} + \frac{\partial^2}{\partial x_{i} \partial x_{j}}
     \right)
\right\}\nonumber
\\
&
\qquad\qquad
+\frac{8}{a^2} \sum_{i=1}^{N-1} \sum_{j=i+1}^{N} \delta^2(x_{i}-x_{j})
\nonumber\\
&= \hat{h}_{4}^{(1)} + 48\kappa^2 \hat{h}_{4}^{(2)} - 4\kappa  \hat{h}_{4}^{(3)} + 8\kappa^2 \hat{h}_{4}^{(4)}
\,\,,
\label{H_4}
\end{align}
where
%
%
$\kappa=\frac{1}{a}$.
%
%


%
For convenience, we introduce the auxiliary operator 
\begin{align}
\hat{Q}_{4} =\frac{1}{\kappa} \hat{H}_{4}
\,\,,
\end {align}
and apply the Hellmann-Feynman theorem to it:
\begin{align}
\langle \chi_N | \left(\frac{d}{d\kappa} \hat{Q}_{4}(\kappa) \right) | \chi_N \rangle
=
\frac{d}{d\kappa} \left(\frac{1}{\kappa} E_{4}(\kappa)\right)
\,\,.
\label{HF}
\end {align}
The left-hand side is related to operator $\hat{H}_{4}$ introduced above by

\begin{widetext}
\begin{align}
\langle \chi_N | \left(\frac{d}{d\kappa} \hat{Q}_{4}(\kappa) \right) | \chi_N \rangle
=
-\frac{1}{\kappa^2} \langle \chi_N | \hat{h}_{4}^{(1)} | \chi_N \rangle
+ 48 \langle \chi_N | \hat{h}_{4}^{(2)} | \chi_N \rangle
+ 8 \langle \chi_N | \hat{h}_{4}^{(4)} | \chi_N \rangle
\,\,.
\label{HF_2}
\end {align}
\end{widetext}

We evaluate the terms of the right-hand side separately. First, we find

%
%
\begin{align}
\begin{split}
&
\langle \chi_N | \hat{h}_{4}^{(1)} | \chi_N \rangle 
=N L \frac{\partial^4}{\partial x^4} \rho_{1}(x;\,0)\Big|_{x=0} 
\\
&
= 12 L n^4 c_{3} \delta(0) + 24 L n^5 c_{4}
\,\,.
\end{split}
\label{h_1_result}
\end {align}

Actually, the infinity in the form of $\delta(0)$ is canceled by the analogous divergence produced by

%
%
\begin{align}
\langle \chi_N | \hat{h}_{4}^{(4)} | \chi_N \rangle=
\frac{1}{2} L n^2 g_2 \delta(0)
\,\,
\label{h_4_result}
\end{align}
as can be shown using the connection of order (3,2),
\begin{align}
\label{conn23}
c_{3}=\frac{1}{3} \frac{1}{(n a)^2}g_2
\,\,.
\end {align}
The latter is deduced from (0,2) above, Eq.(\ref{Eq02}), and (3,0) that reads \cite{Olshanii}
\begin{eqnarray}
c_3=\frac{\gamma^2}{12}e_2'.
\end{eqnarray}
We remark that conversely, starting from the sole requirement that $\hat{H}_4$ is divergence-free, Eq.~(\ref{conn23}) naturally follows from our derivation, which can thus be seen as a new and independent proof of this connection. Then, $(3,0)$ is derived by combination with $(0,2)$.

An other way to derive $(3,2)$ is as follows: due to the contact condition, one can write
\begin{eqnarray}
&&\rho_k(x_1,\dots,x_k;x_1',\dots,x_k')\nonumber\\
&&\!\!\!\!\!\!=\!\!\!\sum_{m=0}^{+\infty}\!\rho_k^{(m)}\!\!\left(\!\frac{x_1\!+\!x_1'}{2},x_2,\dots,x_k;x_2',\dots,x_k'\!\right)\!|x_1\!-\!x_1'|^m\!\!\!.
\end{eqnarray}
Since
\begin{eqnarray}
\rho_1=\int dx_2\rho_2(x_1,x_2;x_1',x_2)
\end{eqnarray}
and
\begin{eqnarray}
&&\int dx_2|x_1\!-\!x_2||x_1'\!-\!x_2|=_{x_1\to x_1'}\frac{1}{3}|x_1\!-\!x_1'|^3+\dots
\end{eqnarray}
where the dots represent a regular function, one finds the general result
\begin{eqnarray}
\rho_k^{(3)}(0,\dots;0,\dots)=\frac{N-k}{3a^2}\rho_{k+1}(0,\dots;0,\dots)
\end{eqnarray}
or, written an other way,
\begin{eqnarray}
c_3^{(k)}=\frac{\gamma^2}{12}g_{k+1},
\end{eqnarray}
a higher-order connection from which $(3,2)$ follows as a corollary.

%
%
To finish with, we evaluate
\begin{align}
\begin{split}
\langle \chi_N | \hat{h}_{4}^{(2)} | \chi_N \rangle 
&
=
\frac{1}{6} L n^3  g_{3}
\,\,.
\end{split}
\label{h_2_result}
\end {align}

%
Inserting (\ref{h_1_result}), (\ref{h_4_result}) and (\ref{h_2_result})
into (\ref{HF}) and (\ref{HF_2}) ends the derivation. We now comment on the physical meaning of Eq.~(\ref{Maineq}). The fact that $g_3$ appears stems from $\hat{h}_4^{(2)}$ in Eq.~(\ref{H_4}), that involves three-body processes provided $N\geq 3$. The coefficient $c_4$, that stems from $\hat{h}_4^{(1)}$, is related to the higher kinetic energy in that the momentum operator applied to the density matrix generates the coefficients of its Taylor expansion when taken at zero distance.

One can even go further, combining the connection (0,3) \cite{Cheianov, Smith},
\begin{align}
\label{g3}
g_3(\gamma)\!=\!\frac{3}{2}\frac{e_4'}{\gamma}\!-\!5\frac{e_4}{\gamma^2}\!+\!\left(1\!+\!\frac{\gamma}{2}\right)\!e_2'\!-\!2\frac{e_2}{\gamma}\!-\!3\frac{e_2e_2'}{\gamma}\!+\!9\frac{e_2^2}{\gamma^2},
\end{align}
with the connection (4,3), Eq.~(\ref{Maineq}), to obtain the connection (4,0),
\begin{align}
\label{c4}
c_4(\gamma)\!=\!\frac{\gamma e_4'}{12}\!-\!\frac{3}{8}e_4\!+\!\frac{2\gamma^2\!+\!\gamma^3}{24}e_2'\!-\!\frac{\gamma e_2}{6}\!-\!\frac{\gamma e_2e_2'}{4}\!+\!\frac{3}{4}e_2^2,
\end{align}
given in \cite{Dunjko} without proof. The right-hand sides of these last two equalities involve moments of the pseudo-momentum distribution only, and it is a general fact that all correlations of the model are defined through connections of type $(l,0)$ and $(0,k)$, as a consequence of integrability.

\section{Conjectures about the exact moments of the density of pseudo-momenta}
\label{Conjecturesmoments}

As illustrated above, local correlation functions are linked to the even-order moments $e_{2k}$ of the density of pseudo-momenta (we recall that odd ones are trivially null by parity). Their exact and explicit analytical expression is not known to date, only the first few terms of their exact asymptotic expansions in the weakly- and strongly-interacting regimes have been computed exactly. To go further and cover the full range of repulsive interaction strengths $\gamma \in [0,+\infty[$, we generalize to arbitrary moments recent conjectures about the ground-state energy.

\subsection{Conjecture in the weakly-interacting regime}

The first conjecture concerns the Taylor expansion of $e_{2k}$ in the weakly-interacting regime, and reads
\begin{align}
\label{conjweak}
e_{2k}(\gamma)=\sum_{i=0}^{+\infty}\frac{a_{2k,i}}{\pi^i}\gamma^{k+i/2},
\end{align}
where $\{a_{2k,i}\}$ are real coefficients. This is a generalization to arbitrary $k$ of the conjecture proposed in \cite{Widom} for $e_2$.

According to Eq.~(\ref{moments}), a trivial necessary condition is $a_{0,i}=\delta_{i,0}$, where $\delta_{.,.}$ is the Kronecker symbol. The exact nontrivial coefficients unambiguously found so far are $a_{2,0}\!=\!1$, $a_{2,1}\!=\!-4/3$ \cite{Lieb}, $a_{2,2}\!=\!\zeta(2)\!-\!1$ \cite{Widom}, where $\zeta$ is the Riemann zeta function, $a_{4,0}\!=\!2$, and $a_{4,1}\!=\!-88/15$ \cite{Cheianov}. Based on accurate numerics by S. Prolhac, G. Lang conjectured $a_{2,3}\!=\!3\zeta(3)/8-1/2$. Prolhac also proposed $a_{2,4}=a_{2,3}/3$ and $a_{2,5}=-45\zeta(5)/1024+15\zeta(3)/256-1/32$ \cite{Prolhac}. Higher-order coefficients $a_{2,i}$ are numerically known with high accuracy up to order $i\!=\!10$ \cite{Prolhac}.

Using the conjecture (\ref{conjweak}) together with the known expansion at low $\alpha$, heuristically introduced in \cite{Hutson} and proven in \cite{Wadati} (we refer to Appendix \ref{refereesugg} for a derivation of the first part),
\begin{align}
&g(z;\alpha)\simeq_{\alpha \ll 1}\frac{\sqrt{1-z^2}}{2\pi\alpha}\nonumber\\
&+\frac{1}{4\pi^2\sqrt{1-z^2}}\left[z\ln\left(\frac{1-z}{1+z}\right)+\ln\left(\frac{16\pi}{\alpha}\right)+1\right],
\end{align}
combined with Eqs.~(\ref{Fredholm}), (\ref{alphagamma}) and (\ref{moments}), we find the general form of the first two coefficients at fixed order $k$,
\begin{align}
a_{2k,0}=\frac{1}{k+1}\binom{2k}{k}= C_k,
\end{align}
where $\{C_k\}$ denote the Catalan numbers, and
\begin{align}
a_{2k,1}\!=\!\binom{2k}{k}\!
-
\frac{2^{4k}}{\binom{2k+1}{k}}\frac{1}{k+1}\sum_{i=0}^k\left[\frac{1}{2^{2i}}\binom{2i}{i}\right]^2
.
\end{align}

\subsection{Conjecture in the strongly-interacting regime}

In the strongly-interacting regime, we generalize a recent conjecture on $e_2$ \cite{Lang2016}, by stating that the asymptotic expansion in $1/\gamma$ is partially resummed in a natural way as
\begin{align}
\label{conjstrong}
e_{2k}(\gamma)=\left(\frac{\gamma}{2+\gamma}\right)^{\!2k}\sum_{i=0}^{+\infty}\frac{\pi^{2(k+i)}}{(2+\gamma)^{3i}}\mathcal{L}_{2k,i}(\gamma)
\end{align}
where $\mathcal{L}_{2k,i}$ are polynomials with rational coefficients, such that $\mathcal{L}_{2k,0}=1/(2k+1)$ and $\mathcal{L}_{2k,i\geq 1}$ is of degree $i-1$. Using a basis of orthogonal polynomials to systematically find a $1/\gamma$ expansion of the moments as explained in \cite{Ristivojevic} and \cite{Lang2016}, together with the conjecture Eq.~(\ref{conjstrong}), we find by identification:
\begin{align}
\label{e4conj}
&\mathcal{L}_{4,1}(X)=\frac{32}{35},\nonumber\\
&\mathcal{L}_{4,2}(X)=-\frac{1984}{1575}X+\frac{3424}{1575},\nonumber\\
&\mathcal{L}_{4,3}(X)=\frac{8192}{3465}X^2-\frac{37376}{5775}X+\frac{169728}{45045},\nonumber\\
&\mathcal{L}_{4,4}(X)=-\frac{47104}{9009}X^3+\frac{59337728}{3378375}X^2\nonumber\\
&-\frac{61582336}{3378375}X+\frac{137573632}{23648625},\nonumber\\
&\mathcal{L}_{4,5}(X)=\frac{192512}{15015}X^4-\frac{765952}{15925}X^3+\frac{80326709248}{1206079875}X^2\nonumber\\
&-\frac{594448384}{14189175}X+\frac{295196160000}{38192529375},\nonumber\\
&\mathcal{L}_{4,6}(X)=-\frac{335872}{9945}X^5+\frac{132872192}{984555}X^4\nonumber\\
&-\frac{2316542492672}{10416144375}X^3+\frac{3689660465152}{18091198125}X^2\nonumber\\
&-\frac{184095784026112}{2406129350625}X+\frac{12238234443776}{1260353469375}.
\end{align}

\section{Illustrations}
\label{Illus}

In this Section, we illustrate the various results obtained above. Within our approach, in order to evaluate $g_3$ and $c_4$ with good accuracy from Eqs.~(\ref{g3}) and (\ref{c4}), it is crucial to correctly evaluate not only $e_2$ and $e_4$, but also their derivatives. For $e_2$, we refer to previous studies \cite{Prolhac, Lang2016}, where it was found that the combination of the weakly-interacting expansion of \cite{Prolhac} and the conjectural partial resummation in the strongly-interacting regime of \cite{Lang2016} yields excellent agreement with accurate numerics. In this work, we checked that $e_2'$ obtained from the conjectures in their respective ranges is also numerically exact for all interaction strengths, as shown in Appendix \ref{e2prime}.

\begin{figure}
\includegraphics[width=8cm, keepaspectratio, angle=0]{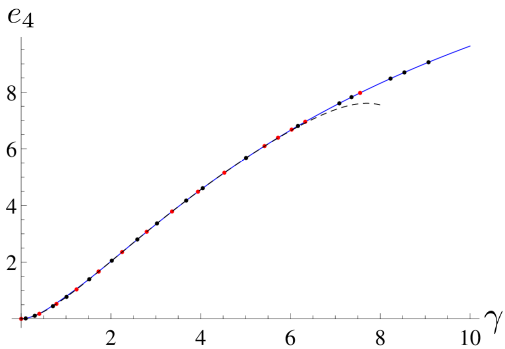}
\caption{(Color online) Dimensionless fourth moment of the distribution of quasi-momenta $e_4$ as a function of the dimensionless interaction strength $\gamma$. Analytical result from the conjecture (\ref{e4conj}) (solid, blue) is in excellent agreement with independent accurate numerics from the authors (red and black dots) for all interaction strengths. The conjecture in the weakly-interacting regime Eq.~(\ref{conjweak}) with appropriate coefficients (black, dashed) reproduces numerical calculations with excellent accuracy up to intermediate interactions.}
\label{Fig1}
\end{figure}

More important, we benchmark the conjectures on the fourth moment. In Fig.~\ref{Fig1} we plot $e_4$ from the conjectures Eq.~(\ref{conjweak}) and Eq.~(\ref{conjstrong}) over the experimentally relevant range $\gamma\in [0,10]$ and find excellent agreement with the numerical integration of the Bethe Ansatz equations (6)-(8). The coefficients used in the weakly-interacting regime are found by fitting data at $\gamma \ll 1$ and the whole curve thereby obtained closely follows numerical solution of the Bethe equations for interaction strengths $\gamma>1$, validating the conjecture. However, our numerical data is not accurate enough to guess the analytical exact value of the unknown coefficients $a_{4,i}$ for $i\geq 2$.

\begin{figure}
\includegraphics[width=8cm, keepaspectratio, angle=0]{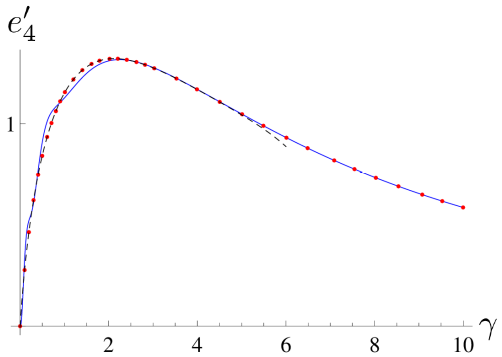}
\caption{(Color online) Derivative of the dimensionless fourth moment of the distribution of quasi-momenta $e_4$ with respect to the dimensionless interaction strength $\gamma$ as a function of the latter. Using the  analytical results  either from the conjectures (\ref{e4conj}) (solid, blue) at strong interactions or  Eq.~(\ref{conjweak}) (dashed, black) at weak interactions, one finds an excellent agreement with accurate numerics (black dots) for all interaction strengths.}
\label{Fig2}
\end{figure}

While no discrepancy between the conjecture from the strongly-interacting regime Eq.~(\ref{conjstrong}) and numerical solution of the Bethe Ansatz equations is seen on this graph, by looking at $e_4'$ shown in Fig.~\ref{Fig2}, one sees that the large-$\gamma$ expansion displays spurious oscillations at intermediate interactions, hence an appropriate combination of both conjectures at weak and strong coupling is needed to recover agreement with numerical calculations over the whole range of interaction strengths.

\begin{figure}
\includegraphics[width=8cm, keepaspectratio, angle=0]{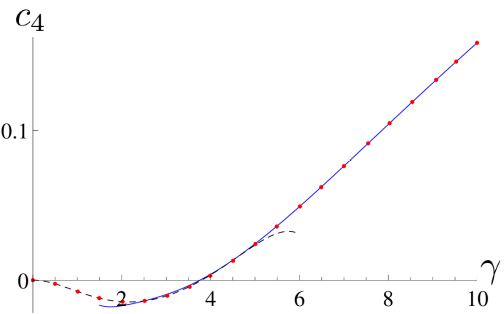}
\caption{(Color online) Dimensionless coefficient $c_4$ as a function of the dimensionless interaction strength $\gamma$, as predicted from the conjectures (solid, blue) and (dashed, black), compared to accurate numerics (red dots). A sign inversion occurs around $\gamma=3.8$.}
\label{Fig3}
\end{figure}

\begin{figure}
\includegraphics[width=8cm, keepaspectratio, angle=0]{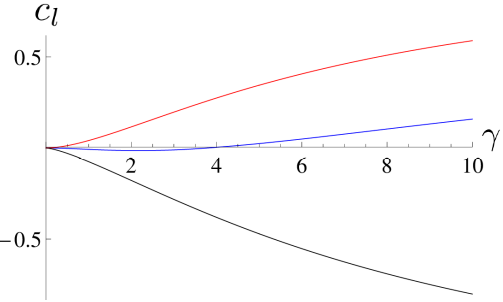}
\caption{(Color online) Dimensionless coefficients $c_2$, $c_4$  and $c_3$  (resp. black, blue, red and from bottom to top) as predicted from conjectures, as functions of the dimensionless interaction strength $\gamma$.}
\label{Fig4}
\end{figure}

We have also checked that our numerical data for $e_2$ and $e_4$, when used in Eq.~(\ref{g3}), yield $g_3(\gamma)$ in close agreement with accurate approximate expressions obtained in \cite{Cheianov} by fitting on the numerical solution of Eqs.~(\ref{Fredholm}), (\ref{alphagamma}), (\ref{moments}) and (\ref{g3}), as illustrated in Appendix \ref{g3gamma}. Having performed all these verifications, we plot in Fig.~\ref{Fig3} the coefficient $c_4$ as a function of $\gamma$ from numerical calculations and the conjectures on $e_2$ and $e_4$. We find that $c_4$ changes sign at $\gamma=\gamma_c\simeq 3.8$. We evaluated the value with more accuracy as $\gamma_c=3.8160616255908\dots$ by comparing two independent numerical solutions of the Bethe Ansatz equations, with agreement of all digits up to this order. This change of sign had already been predicted, based on numerical analysis in \cite{Caux}, that suggested $1<\gamma_c<8$. In Fig.~\ref{Fig4} we plot the known coefficients $c_2$, $c_3$ and $c_4$ as functions of $\gamma$. They are also known by direct calculation in the Tonks-Girardeau regime of infinite interaction strength, where their values are $c_2\!=\!-\pi^2/6$, $c_3\!=\!\pi^2/9$ and $c_4\!=\!\pi^4/120$ respectively, in agreement with our results in the limit $\gamma \to +\infty$, but higher-order terms are known as well in this regime, such as $c_5=-11\pi^4/1350 \dots$ \cite{Tracy, Gangardt, Forrester}. Note that very high interaction strengths are needed to approach the value in the Tonks-Girardeau regime up to a few percents, thus it is quite difficult, for the observable $\rho_1$, to reach this regime experimentally.

\section{Conclusions and outlook}
\label{Outlook}

In conclusion, in this work we have derived an exact relation linking the fourth coefficient of the Taylor expansion of the one-body correlation function at short distances and the local three-body correlation function of the Lieb-Liniger model. This connection can be recast in a form where $c_4$ is expressed in terms of moments of the density of pseudo-momenta. We have investigated the fourth moment $e_4$ in detail and provided new conjectural expressions that are extremely accurate in the whole range of interaction strengths. Both analytically and numerically, we find that $c_4$ changes sign around $\gamma=3.8$.

In outlook, it would be interesting to investigate the link between $c_5$ and the coefficient of the first subleading, high-momentum $1/p^6$ term of the momentum distribution $n(p)$ of the gas, beyond the well-known Tan contact (i.e. the coefficient of the leading $1/p^4$ term). Knowing more terms of the Taylor expansion of $g_1$ and $n(p)$ also allows to probe in a finer way the validity of the Renormalization Group-Luttinger liquid approach by comparing their predictions at large distances or short momenta \cite{Dunjko}. In the perspective of taking an harmonic trapping into account, it also allows to discuss the validity of the Local Density Approximation \cite{Olshanii} by comparison with exact numerics \cite{Decamp}. To this aim, the equation of state from the $1/\gamma$ expansion is accurate in the strongly-interacting regime, while the conjecture thereby deduced is also accurate at intermediate $\gamma$ \cite{Minguzzi}. The attractive regime of the super-Tonks-Girardeau gas may also provide new insights \cite{Trombettonibis, Piroli}. To finish with, the equivalence between Eq.~(\ref{g3}) and an other one derived in \cite{Kormos, Poszgay}, that does not involve the momenta $e_{2k}$ but requires solving other Fredholm integral equations instead, checked numerically so far, still awaits rigorous proof. It shall provide an interesting alternative way to tackle connections, especially in out-of-equilibrium situations. An other approach from field theory, based on an appropriate non-relativistic limit of the sinh-Gordon model, has also shown remarkable efficiency already \cite{Mussardo, Trombettoni} as compared with previous Bethe Ansatz results \cite{Kheruntsyan, Shlyapnikov, Ganshlya}. The full characterization of the local correlations $g_k$ seems, however, especially challenging and insightful. Even the connection (0,3) has not been derived yet based on Bethe Ansatz techniques only.

\acknowledgments

We acknowledge financial support from the ANR SuperRing (ANR-15-CE30-0012-02), and National Science Foundation grants PHY-1402249 and PHY-1607221. We thank the referee for suggesting reference \cite{Fateev}.

\appendix

\section{Derivation of the semi-circular law}
\label{refereesugg}
In this Appendix, we provide a new derivation of the semi-circular law verified by the density of pseudo-momenta at very small interactions, i.e. the dominant term of the dominant term of the solution to Eq.~(\ref{Fredholm}) in the limit $\alpha\to 0$, that reads
\begin{eqnarray}
\label{SC}
g(z;\alpha)\simeq_{\alpha\ll 1}\frac{\sqrt{1-z^2}}{2\pi\alpha}.
\end{eqnarray}

This derivation is based on the methods of Ref.~\cite{Fateev}. First, the kernel of the integral equation is rewritten as
\begin{eqnarray}
\frac{2\alpha}{\alpha^2+(y-z)^2}=\int_{-\infty}^{+\infty}dp\quad\!\!\!e^{-\alpha |p|}\cos[p(y-z)].
\end{eqnarray}
Then, the Fourier integral theorem, stating that
\begin{eqnarray}
f(z)=\frac{1}{2\pi}\int_{-\infty}^{+\infty}dy f(y)\int_{-\infty}^{+\infty}dp \cos[p(y-z)],
\end{eqnarray}
is applied to $g$ to transform Eq.~(\ref{Fredholm}) into
\begin{eqnarray}
\label{EqLiebtrans}
\!\!\!\!\!2\!\int_{0}^{+\infty}\!\!\!\!\!\!dp \left(1\!-\!e^{-\alpha p}\right)\!\!\int_{-1}^1\!\!dyg(y;\alpha)\cos(py)\!=\!1
\end{eqnarray}
after a few simplifications.

To go further, $g(y;\alpha)$ is expanded over the basis of Gegenbauer polynomials $\{U_n\}$, that are orthogonal on $[-1,1]$, as
\begin{eqnarray}
g(y;\alpha)=\sqrt{1-y^2}\sum_{n=0}^{+\infty}A_{n}(\alpha)U_{2n}(y).
\end{eqnarray}
Thus, to lowest order in $\alpha$, Eq.~(\ref{EqLiebtrans}) becomes
\begin{eqnarray}
\label{EqLiebtransapprox}
2\alpha \int_{0}^{+\infty}dp\quad \!\!\!\!p\int_{-1}^1dy \sqrt{1-y^2}A_0\cos(py)=1,
\end{eqnarray}
and using the property
\begin{eqnarray}
\int_{-1}^1dy \sqrt{1-y^2}\cos(py)=\pi\frac{J_1(p)}{p},
\end{eqnarray}
where $J_1$ is the first Bessel function, Eq.~(\ref{EqLiebtransapprox}) becomes
\begin{eqnarray}
2\pi A_0 \alpha \int_{0}^{+\infty}dp J_1(p)=2\pi A_0=1,
\end{eqnarray}
and one obtains Eq.~(\ref{SC}) as announced.

\section{Test of the conjectures for $e_2'$}
\label{e2prime}
\begin{figure}
\includegraphics[width=8cm, keepaspectratio, angle=0]{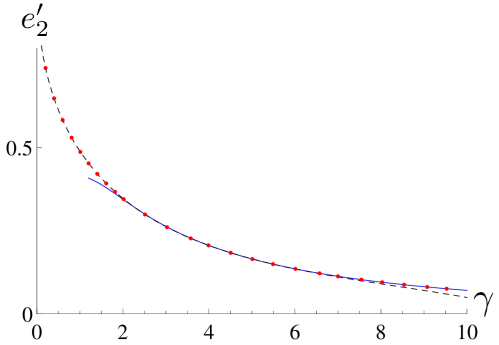}
\caption{(Color online) Derivative $e_2'$ of the dimensionless energy as a function of the dimensionless interaction strength $\gamma$, as predicted from the conjectures (solid, blue) and (dashed, black), compared to accurate numerics (red dots).}
\label{Fig5}
\end{figure}

In Fig.~\ref{Fig5}, we plot $e_2'$ from the conjectures with known coefficients from \cite{Prolhac, Lang2016} and compare with accurate numerics. The agreement is pretty satisfying over the whole range of interactions.

\section{Test of our numerical data for $g_3$}
\label{g3gamma}

\begin{figure}
\includegraphics[width=8cm, keepaspectratio, angle=0]{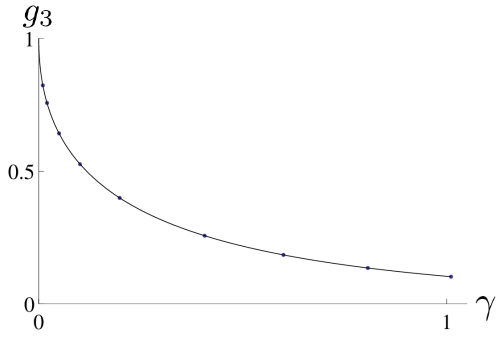}
\includegraphics[width=8cm, keepaspectratio, angle=0]{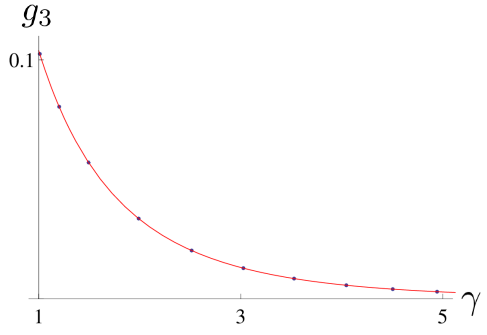}
\includegraphics[width=8cm, keepaspectratio, angle=0]{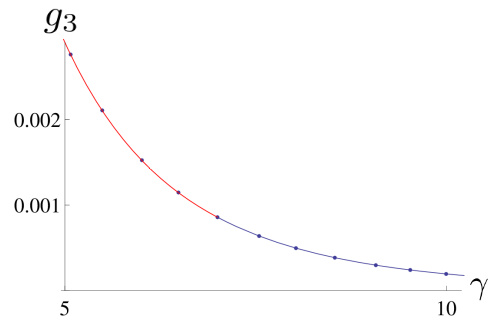}
\caption{(Color online) Three-body local operator $g_3$ as a function of the dimensionless interaction strength $\gamma$ obtained numerically (dots) compared to the three accurate expressions in \cite{Cheianov} (solid lines, black, red and blue) in a wide range of interaction strengths.}
\label{Fig6}
\end{figure}

In Fig.~\ref{Fig6}, we compare the local three-body operator $g_3$ obtained numerically and accurate expressions from Ref.~\cite{Cheianov}. The agreement is excellent.

\end{document}